\documentclass[aps,prl,showpacs,floatfix,twocolumn,amsmath,amssymb,superscriptaddress]{revtex4-1}
\usepackage{graphicx}
\usepackage{color}
\usepackage{ulem}
\usepackage{bm}
\usepackage{hyperref}
\usepackage{bm}
\usepackage{amsfonts}
\usepackage{amssymb}
\usepackage{amsmath}
\usepackage{dsfont}
\usepackage{hyperref}
\usepackage{graphicx}
\usepackage{multirow}
\usepackage{color}

\begin{document}
\title{Unified quantitative model for magnetic and electronic spectra of the undoped cuprates}

\author{B. Dalla Piazza}
\affiliation{Laboratory for Quantum Magnetism, \'Ecole Polytechnique F\'ed\'erale de Lausanne (EPFL), CH-1015, Switzerland}

\author{M. Mourigal}
\affiliation{Laboratory for Quantum Magnetism, \'Ecole Polytechnique F\'ed\'erale de Lausanne (EPFL), CH-1015, Switzerland}
\affiliation{Institut Laue-Langevin, BP 156, 38042 Grenoble Cedex 9, France}

\author{M. Guarise}
\affiliation{Laboratory of Photoelectron Spectroscopy, \'Ecole Polytechnique F\'ed\'erale de Lausanne (EPFL), CH-1015, Switzerland}

\author{H. Berger}
\affiliation{Laboratory of Photoelectron Spectroscopy, \'Ecole Polytechnique F\'ed\'erale de Lausanne (EPFL), CH-1015, Switzerland}

\author{T. Schmitt}
\affiliation{Swiss Light Source, Paul Scherrer Institut, CH-5232 Villigen PSI, Switzerland}

\author{M. Grioni}
\affiliation{Laboratory of Photoelectron Spectroscopy, \'Ecole Polytechnique F\'ed\'erale de Lausanne (EPFL), CH-1015, Switzerland}

\author{H. M. R\o{}nnow}
\affiliation{Laboratory for Quantum Magnetism, \'Ecole Polytechnique F\'ed\'erale de Lausanne (EPFL), CH-1015, Switzerland}

\date{\today}

\begin{abstract}
Using low-energy projection of the one-band $t$-$t'$-$t''$-Hubbard model we derive an effective spin-Hamiltonian
and its spin-wave expansion to order $1/S$. We fit the spin-wave dispersion of several parent compounds to the high-temperature superconducting cuprates: La$_2$CuO$_4$,
Sr$_2$CuO$_2$Cl$_2$ and Bi$_2$Sr$_2$YCu$_2$O$_8$. Our accurate
quantitative determination of the one-band Hubbard model parameters allows prediction and comparison to experimental
results of measurable quantities such as staggered moment, double occupancy density, spin-wave velocity and bimagnon
excitation spectrum and density of states, which is discussed in relation to K-edge RIXS and Raman experiments.
\end{abstract}


\maketitle


High-$T_c$ superconductors challenge all known theoretical approaches 
by mixing charge and magnetic degrees of freedom and lacking a small 
variational parameter. The One-Band Hubbard Model (1bHub), proposed by 
Anderson~\cite{anderson_resonating_1987} to describe their CuO$_2$ 
planes, includes both of these difficulties in its 
parameters, the electron filling, the 
hopping matrix element $t$ and the Coulomb repulsion $U$. The ratio $t/U$
is moderately small in the cuprates, thus most approaches start by projecting out 
double occupancies (DO) to obtain the Heisenberg model at
half-filling or the $t$-$J$ model 
for hole- or electron-doping. Such a projection
is too complicated to be carried out exactly but may be performed as 
an expansion in powers of $t/U$~\cite{macdonald_t/u_1988}. 

Experimentally, different techniques have probed separate channels -- magnetic or electronic. Inelastic Neutron 
Scattering (INS) on La$_2$CuO$_4$~\cite{coldea_spin_2001} demonstrated
that the  projection must be carried out at least to the fourth order ($t^4/U^3$)
to reproduce the observed magnetic 
excitation spectrum. Angle-Resolved Photo-Emission 
Spectroscopy (ARPES) indicates that first, second and third nearest 
neighbor hopping matrix elements are needed to reproduce the observed
electronic quasiparticle dispersion~\cite{RevModPhys.75.473}. A 
quantitative description of the undoped high-$T_c$ parent compounds
therefore needs a model incorporating both those 
considerations~\cite{delannoy_low-energy_2009}.

In this letter, we develop such a quantitative theory and present 
the resulting sets of 1bHub\ parameters for single and bilayer undoped cuprates 
Sr$_2$CuO$_2$Cl$_2$~\cite{PhysRevLett.105.157006}, 
La$_2$CuO$_4$~\cite{PhysRevLett.105.247001} and Bi$_2$Sr$_2$YCu$_2$O$_8$
[this work], obtained
by fitting their magnetic excitation spectra measured by Resonant 
Inelastic X-ray Scattering (RIXS) and INS.\\
\begin{figure}
\includegraphics[width=0.90\linewidth]{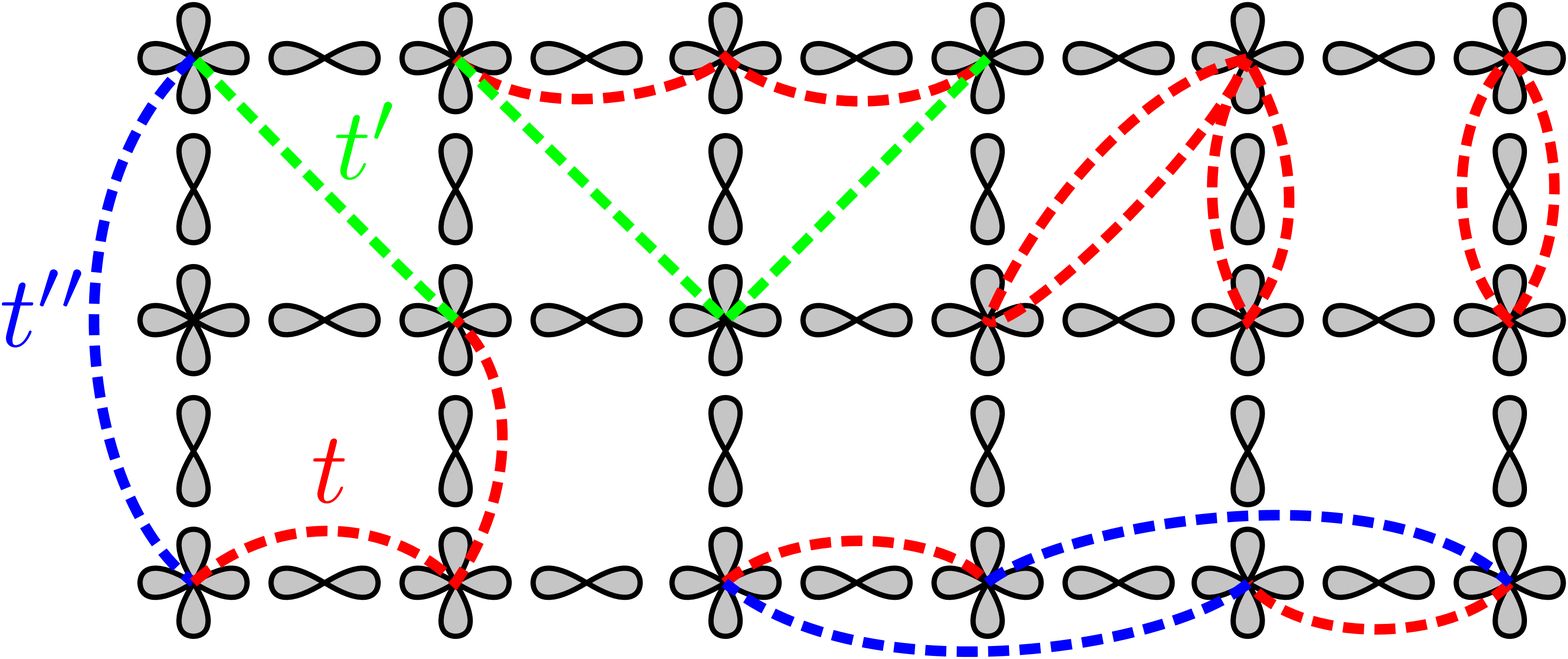}
	\caption{\label{hoppings} (Color Online)
	CuO$_2$ planes from the perovkite 
	structure. Examples of exchange loops from the effective spin 
	Hamiltonian of Eq.~\ref{effective_spin_ham} are indicated with first,
	second and third nearest neighbor hoppings $t$, $t'$ and $t''$.}
\end{figure}
%
%
We start from the 1bHub\ Hamiltonian given by
	\begin{equation} 
		\hat{\mathcal{H}}=-\sum_{i,j,\sigma}t_{ij}c_{i,\sigma}^\dagger 		
		c_{j,\sigma}+U\sum_{i}n_{i,\uparrow}n_{i,\downarrow}
	\end{equation}
where $c^\dagger_{i,\sigma}$ and $c_{i,\sigma}$ creates or destroys a 
fermion with spin $\sigma$ on site $i$, $t_{ij}$ is
the hopping matrix element between sites $i$ and $j$, $U$ the effective on-site 
repulsion and $n_{i,\sigma}\!=\!c^\dagger_{i,\sigma}c_{i,\sigma}$ the
fermionic number operator. 
At half filling, the kinetic term mixes states with different number of DO, which in the limit $t/U\ll1$ separate into different energy scales.
We use the unitary transformation 
technique~\cite{macdonald_t/u_1988} up to order $t^4/U^3$
to decouple these states and obtain the 
effective spin Hamiltonian
	\begin{eqnarray}\label{effective_spin_ham}
	\displaystyle{\hat{\mathcal{H}}^{(4)}} &=&
	\displaystyle{\sum_{\left\{
		\begin{minipage}{0.04\linewidth}
			\includegraphics[width=\linewidth]{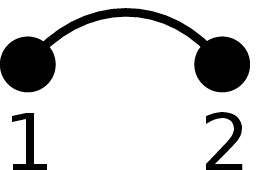}
		\end{minipage} 
	\right\}}}
	 \left(\frac{4t_{12}^2}{U}-\frac{16t_{12}^4}{U^3}\right)
	 \mathbf{S}_1 \mathbf{S}_2
	\hspace{0.1cm}+\hspace{-0.1cm}\displaystyle{\sum_{\left\{
		\begin{minipage}{0.04\linewidth}
			\includegraphics[width=\linewidth]{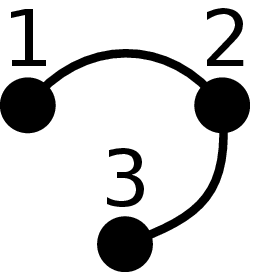}
		\end{minipage}\right\}}} 
	\frac{4t_{12}^2t_{23}^2}{U^3}
	\mathbf{S}_1 \mathbf{S}_3 \nonumber \\*
	&-&\displaystyle{\sum_{\left\{
		\begin{minipage}{0.045\linewidth}
			\includegraphics[width=\linewidth]{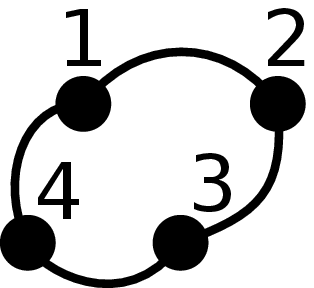}
		\end{minipage}
		\right\}}}
	\frac{4t_{12}t_{23}t_{34}t_{41}}{U^3}\Big\{
	\sum_{\begin{subarray}{c}i,j=1\\i\neq j\end{subarray}}^4
		\mathbf{S}_i \mathbf{S}_j
		- 20\Big[
		\left(\mathbf{S}_1 \mathbf{S}_2\right)
		\left(\mathbf{S}_3\mathbf{S}_4\right)
	  	\nonumber \\*
	&&+\left(\mathbf{S}_1 \mathbf{S}_4\right)
		 \left(\mathbf{S}_2\mathbf{S}_3\right)
		-\left(\mathbf{S}_1 \mathbf{S}_3\right)
		 \left(\mathbf{S}_2\mathbf{S}_4\right) \Big]\Big\} + {E^{(4)}},
	 \end{eqnarray}
where ${E^{(4)}}$ is a constant and
$\left\{\begin{minipage}{0.06\linewidth}
	\includegraphics[width=\linewidth]{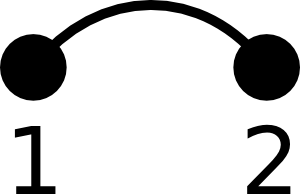}
	\end{minipage}\right\}$,
$\left\{\begin{minipage}{0.06\linewidth}
	\includegraphics[width=\linewidth]{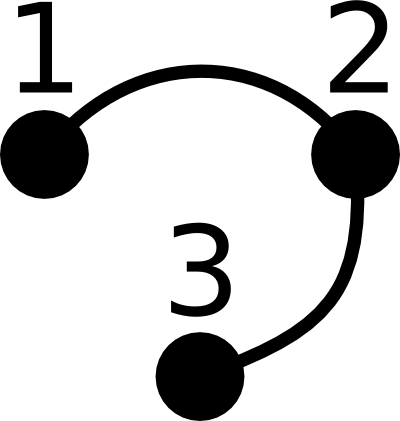}\end{minipage}\right\}$ 
and $\left\{\begin{minipage}{0.07\linewidth}
	\includegraphics[width=\linewidth]{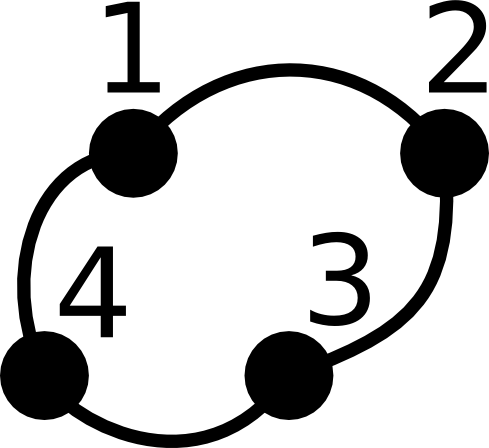}
\end{minipage}\right\}$ stand for the plaquette ensembles of two, 
three and four sites respectively, and connected as sketched. We emphasize 
that the loop ensembles are fully defined by the considered 
hopping matrix elements $t_{ij}$ and the lattice topology, so that the 
Hamiltonian of Eq.~\ref{effective_spin_ham} is the general low-energy 
projection to order $t^4/U^3$ of the 1bHub\ at half filling for any lattice.
Examples of plaquettes of two, three and four sites 
involving first, second and third Nearest Neighbor (NN) hopping amplitudes $t$, 
$t'$ and $t''$ are sketched on Fig.~\ref{hoppings}. A similar development but 
away from half-filling would result in a $t$-$J$ model with the same 
magnetic couplings as in Eq.~\ref{effective_spin_ham} plus a family of charge
plaquette hoppings.
Being the canonical model for high-$T_c$ superconductors,
a large variety of analytic and numerical approaches exist to study such a model
\cite{ISI:000236516300002}.

%
%
In contrast to the doped compounds, the derivation of measurable 
quantities is much easier for their undoped parents. 
Here, we use Spin-Wave Theory (SWT) to derive the  
dispersion of the magnetic excitations which are measured 
by INS or RIXS. We expand
the spin operators of Eq.~\ref{effective_spin_ham} in terms of 
Holstein-Primakov bosons. Keeping the first $1/S$ correction
for the $t_{ij}^2/U$ terms, the Hamiltonian transforms as $\hat{\mathcal{H}}^{(4)} = E_N + 
\hat{\mathcal{H}}_2 + \hat{\mathcal{H}}_4 + {\mathcal{O}(1/S)}$,
where the N\'eel ground-state energy $E_{\rm N}$ includes the constant of 
Eq.~{\ref{effective_spin_ham}}, 
and the harmonic dispersion \ensuremath{\omega_{0}({\bf k})}\ is obtained from a Bogoliubov transformation 
of the quadratic term $\hat{\mathcal{H}}_2$. A Hartree-Fock 
decoupling~\cite{Oguchi1960} of the quartic term $\hat{\mathcal{H}}_4$ results 
in an overall momentum-dependent
correction to the magnon energy so that our final spin-wave Hamiltonian reads
	\begin{eqnarray}\label{1S_ham}
		\hat{\mathcal H}^{(4)} &=& 
		\sum_{\bf k} Z_c({\bf k})
		\ensuremath{\omega_{0}({\bf k})} \alpha_{\bf k}^\dagger\alpha_{\bf k} + E_{\rm N} + \delta E
	\end{eqnarray}
where $\alpha$'s are free magnon 
operators, $Z_c(\bf k)$ is the $1/S$ renormalization of their dispersion and 
$\delta E$ is the quantum correction to the ground state energy at this 
order. For the bilayer square-lattice, Eq.~\ref{effective_spin_ham} is 
still valid but the loop ensembles now include interlayer hopping
$t_{\perp}$, and two boson flavors to account for the top 
and bottom sites. This results in two magnon modes which are  
gapped respectively at $(0,0)$ and $(\pi,\pi)$ but degenerate along the magnetic 
Zone Boundary (ZB).


\begin{figure}
\includegraphics[width=\linewidth]{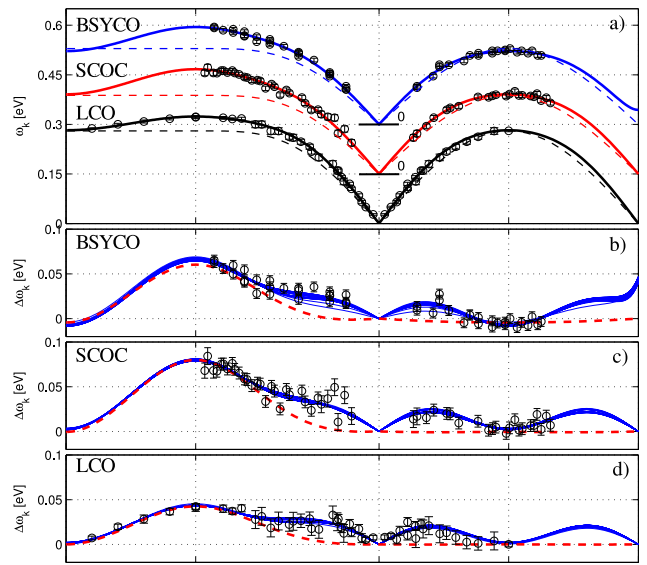}
\caption{\label{swfits} (Color Online) 
(a) Fits of the measured SW dispersions (solid lines) and the
1JHei{} SW dispersion (dashed lines) with $J_{\rm NN}=\omega_{(\pi/2,\pi/2)}/2Z_c$. From top to bottom: 
BSCCO ($J_{\rm NN}\!=\!0.15$~eV), SCOC~\cite{PhysRevLett.105.157006} ($J_{\rm NN}\!=\!0.12$~eV) and
LCO~\cite{PhysRevLett.105.247001} ($J_{\rm NN}\!=\!0.14$~eV). (b-d) Same as in (a) but
subtracted from the 1JHei{} SW dispersion to enhance details,
BSCCO, SCOC and LCO respectively. Data points from experiments
are folded onto the equivalent high-symmetry axes of the first Brillouin zone.}
\end{figure}
%
%
Before applying the result to the
cuprate compounds SCOC~\cite{PhysRevLett.105.157006} and
LCO~\cite{PhysRevLett.105.247001}, we report
here new measurements of the spin-wave (SW) dispersion in 
Bi$_2$Sr$_2$YCu$_2$O$_8$, a bilayer parent compound.
Single crystals were grown by the
flux method with yttrium ensuring an insulating 
antiferromagnetic phase. The SW dispersion was measured using Cu $L_3$ edge RIXS at 
the SAXES end-station of the Swiss Light Source ADRESS beamline, experimental details 
and data analysis as described previously \cite{PhysRevLett.105.157006}.

\begin{figure*}[t!]
\includegraphics[width=\linewidth]{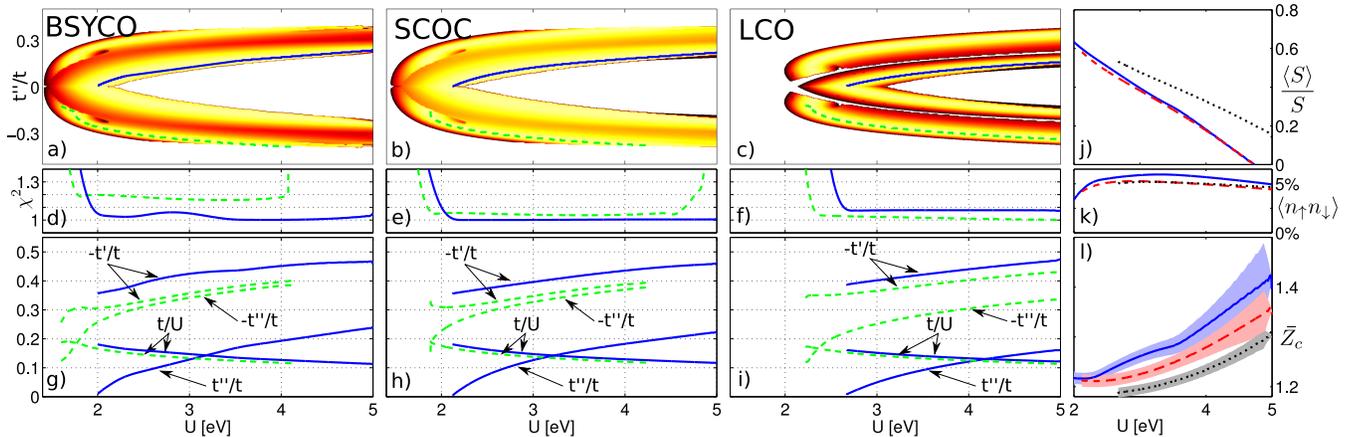}
\caption{\label{swfitgood}(Color Online) Summary of the fitting results. 
(a-c) Goodness ($\chi^2$) of the $(t/U,t'/t)$ fit 
for fixed $(U,t''/t)$ in 
the case Bi$_2$Sr$_2$CaCu$_2$O$_8$, Sr$_2$CuO$_2$Cl$_2$ and La$_2$CuO$_4$
respectively. Solid (blue) and dashed (green) lines are 
best fit solutions as function of $U$ for $t't''<0$ and $t't''>0$ 
respectively. (d-f) $\chi^2$ along the best fit lines defined in (a-c). 
(g-i) $(U,t,t',t'')$ solutions along the best fit lines defined in (a-c).
(j-l) Staggered magnetization, DO density and $\bf k$-averaged
quantum $1/S$ renormalization $Z_c$ along the $t't''<0$ best fit lines defined in (a-d).
Lines: Solid blue BSYCO, dashed red SCOC and dotted black LCO.
Shaded areas indicate $\bf k$-variation of $Z_c$}
\end{figure*}
%
%
The spin-wave dispersions of the various compounds are shown in 
Fig.~\ref{swfits}. They all feature a dispersion between the
ZB points $(\pi,0)$ and $(\pi/2,\pi/2)$ which can in principle be 
explained by the effective model of Eq.~\ref{effective_spin_ham} 
with NN hopping alone (dashed red lines). However, this approach 
results in unphysically low $U\!=\!2.2$~eV~\cite{coldea_spin_2001} for LCO and 
$U\!<\!2$~eV for SCOC and BSYCO. From the former to the latter, their 
ZB dispersions respectively reach $40$, $70$ and $55$~meV. Although $U$ is an
effective on-site repulsion, closer to the charge-transfer gap 
than to the bare Coulomb repulsion, a good 1bHub\ must use an effective 
parameter compatible with electronic and optical spectroscopies, which 
request $U\!\sim\!3$-4~eV for the cuprates~\cite{ISI:000086788300001}.
%
%
Consistently with ARPES results, we therefore include second and third NN  hopping in our effective 
model and derive the spin-wave dispersion of Eq.~\ref{1S_ham} which is now a 
function of four parameters $(U,t,t',t'')$. The measured SW 
dispersions contain three distinct constraints, the $(\pi,0)$ and 
$(\pi/2,\pi/2)$ ZB energies and the spin-wave velocity. We thus expect a 
one-dimensional solution and choose the free parameter to be the effective 
$U$.
%
%
The fitting procedure is as follows. For a fixed choice of $(U,t''/t)$ 
we start by fitting the two other parameters $t/U$ and $t'/t$. 
As the calculation of 1/S estimate of 
$Z_c$ involve a slowly convergent integration over $\bf k$-space, we include 
it in a two-step iterative approach. First, we fix its value to the uniform 
$Z_{0}\!=\!1.1579$ obtained for the NN Heisenberg model (1JHei). 
Then, we fit $Z_{0}\omega_{\bf 
k}(U,t/U,t'/t,t''/t)$ using a non-linear least-squares algorithm and 
calculate a first non-uniform $Z_{1}(\bf k)$ from the obtained 
parameters set. We iterate this procedure until $Z_{n}(\bf k)$ 
converges, typically after 10-15 steps. In the case of BSYCO, 
we further include an interplane hopping $t_\perp$. However, the 
resolution of RIXS does not allow to distinguish the splitting between 
the two magnon modes and we fix it to the value
$t_\perp\!=\!54$~meV reported by Chuang~\textit{et al.}~\cite{PhysRevB.69.094515}.
%
%

The fitting results over the $(U,t''/t)$ plane are shown in 
Fig.~\ref{swfitgood} with the $(U,t/U,t'/t,t''/t)$ parameters along the best fit lines.
Overall, the four compounds share common features 
\textit{i.e.}\ a strong lower boundary for $U$, an increase of $|t'|/t$ and $t''/t$ with 
$U$, and a slowly varying $t/U$. Due to the 
$t^2t't''/U^3$ term, the calculated magnon dispersion is symmetric in the 
signs of $t$ and $t'$ but not in the relative sign of $t'$ and $t''$ 
resulting in two separate solutions for $t't''\!<\!0$ and $t't''\!>\!0$. From 
the best-fit lines, one can see that the inclusion of $|t''|$ is necessary in 
order to get $U\approx3$-$4$~eV.
For some regions of the $(U,t''/t)$ space, the 
N\'eel state is not the classical ground state of Eq.~\ref{effective_spin_ham}, 
and/or it is destroyed by quantum fluctuations. Both cases can be 
systematically 
determined by looking at the size of zero-point fluctuations $\langle 
a_i^\dagger a_i\rangle$. The outermost void regions in Fig.~\ref{swfitgood}(a)-(d) 
are those where N\'eel order is unstable. In Fig.~\ref{swfitgood}(m), we 
calculate the evolution of the staggered magnetization as a function of 
$U$ along the best fit lines. Increasing $U$, 
$t'$ and $t''$ grow while $t/U$ stays roughly constant, bringing more 
frustration and subsequently reducing the ordered moment. 
We calculate the double occupation density using the Feynman-Hellmann theorem
$\langle n_{i,\uparrow}n_{i,\downarrow}\rangle=\partial\langle\hat{\mathcal{H}}^{(4)}\rangle/\partial U$.
Along the best fit lines, a $U$-independent value $5\%$ is found for all cuprates in agreement
with the electronic shielding factor calculated in ref.~\cite{PhysRevB.72.224511}.
The $1/S$ estimate of $Z_c(\bf k)$ is found to vary only
about $2\%$ across the Brillouin zone. In Fig.~\ref{swfitgood}(o) we show
the average value, which vary from $1.2$ to $1.3$ for $U\approx 3$-$4$~eV
which again reveals the prominent role of quantum fluctuations in the range of 
parameters relevant for the cuprates.
%
%

The effective $U$ cannot be directly obtained
through magnetic excitations. However, more direct experimental techniques may
give good estimates of $U$ thus determining a unique set of
1bHub\ parameters for each of the above compounds. In particular,
an estimate of the 1bHub\ parameters of SCOC was obtained from 
ARPES~\cite{ISI:000086788300001} as $U\!=\!3.5$~eV, $t\!=\!0.35$~eV, 
$t'\!=\!-0.12$~eV and $t''\!=\!0.08$~eV. Consistently with those parameters and in order to compare
the three cuprate compounds, we adopt in Tab.~\ref{params} a uniform value
$U\!=\!3.5$~eV and the $t't''\!<0\!$ solution. A more accurate determination of $U$ could be found 
in the charge-transfer (CT) excitation part of the RIXS spectrum. Using the 
above ARPES parameter estimates, Hasan~\textit{et al.}\ could identify
a dispersing excitation around $3$~eV in Cu K-edge RIXS as CT excitation~\cite{Hasan09062000}. A 
similar approach using our parameter sets would allow unambiguous 
determination of $U$.
\begin{table}[t!]
\begin{tabular}{c|c|c|c|c|c|c}
		&$t$  [meV]	&$t'$ [meV]	&$t''$[meV]	&$\langle S \rangle/S$	&$c$ [meV\AA]	&$\langle n_{i,\uparrow}n_{i,\downarrow}\rangle$\\\hline
BSYCO	&$470(10)$	&$-205(3)$	&$79(4)$	&$0.3$					&$0.146$		&$5.9\%$										\\\hline
SCOC	&$480(10)$	&$-200(5)$	&$75(5)$	&$0.29$					&$0.163$		&$5.1\%$										\\\hline
LCO		&$492(7)$	&$-207(3)$	&$45(2)$	&$0.4$					&$0.195$		&$5.2\%$										\\
\end{tabular}
\caption{ \label{params} Parameters determined from the spin-wave fits with fixed $U\!=\!3.5$~eV and $t't''\!<\!0$.}
\end{table}
%
%

\begin{figure}
\includegraphics[width=\linewidth]{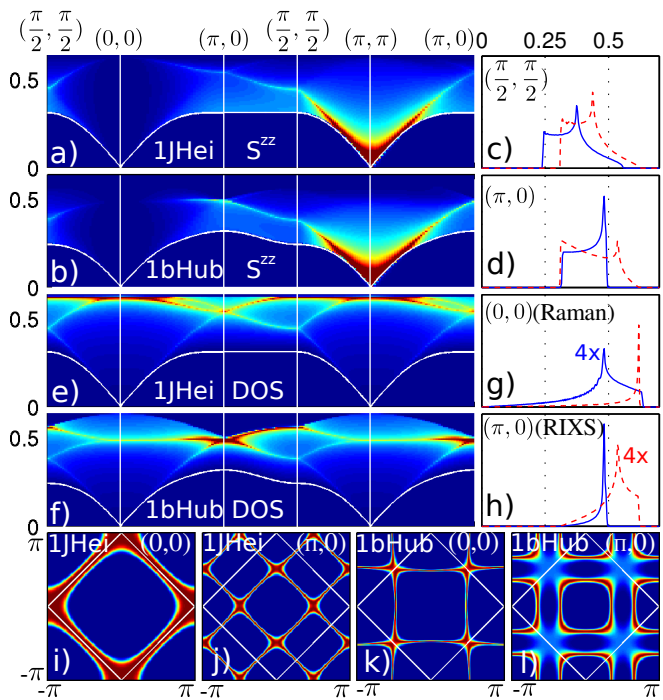}
\caption{\label{stag_magn}(Color Online) (a-b) 
$S^{zz}(\bf k,\omega)$ for 1JHei\ with $J_{\rm NN}=\omega_{(\pi,0)}/2Z_c$ and the 1bHub\ with the SCOC 
parameters. (c-d) $S^{zz}$ energy lineshapes at the ZB 
points for 1JHei (dashed red line) and 1bHub (solid blue line). 
(e-f) Corresponding two-magnon DOS. (g-h) DOS 
profiles at $(0,0)$ and $(\pi,0)$. (i-j) Single magnons wave-vectors contributing to the 
DOS peaks at $(0,0)$ and $(\pi,0)$ in 1JHei, (k-l) in the SCOC 1bHub.}
\end{figure}

Having established a quantitative model for the SW dispersion allows to
predict further quantities. We compute the non-interacting two-magnon dynamical structure 
factor $S^{zz}(\bf k,\omega)$ probed by INS, and the two-magnon density of 
states (DOS) underlying, at $(0,0)$, Raman scattering and, at $(\pi,0)$, K-edge RIXS~\cite{PhysRevB.81.085124}.
Although higher-order magnon interaction affect those two-magnon
quantities~\cite{Canali1993,Canali1992,Nagao2007} our results already allow several observations.
Compared to 1JHei, our predictions show the  enhancement of a $500$~meV peak in $S^{zz}$ at $(\pi,0)$~[Fig.~4(d)]
which shows that attempts
to explain the reported INS lineshape at $(\pi,0)$ from quantum effects must
consider the full Hamiltonian presented here~\cite{christensen_quantum_2007,PhysRevLett.105.247001,Tsyrulin2010}.
Also, along the ZB,
the intensity of the one-magnon (transverse) excitations is 
constant so that the missing spin-wave amplitude observed by INS 
~\cite{PhysRevLett.87.037202,Christensen2004,christensen_quantum_2007,PhysRevLett.105.247001} \textit{does not} 
result from further neighbor hopping.\\
In the 1JHei, the (0,0) two-magnon DOS peaks at $4Z_cJ_{\rm NN}$, corresponding to creating
two spin-waves at the ZB. The peak in Raman B$_{1g}$ spectra is found
at 0.37 eV for SCOC~\cite{PhysRevB.53.R11930} corresponding to $\sim 2.8 J_{\rm NN}$. The reduced energy was explained as due to
magnon-magnon interactions~\cite{Canali1992}, but the peak-width could not be reproduced. The
large ZB dispersion that our model entails firstly imply that experiments should not be compared
to a single $J_{\rm NN}$, secondly it explains the Raman peak width as a
range of energies from $2\omega_{(\pi,0)}$ extending down to $2\omega_{(\frac{\pi}{2},\frac{\pi}{2})}$,
where a maximum occurs because at this energy entire lines in one-magnon momentum
space contribute [Fig.~4(k)]. Thirdly, it predicts a lower peak energy requiring weaker magnon interaction to match experiments.
For a correct calculation of magnon interactions, we caution that in the cuprates, it is different one-magnon
that contribute to the Raman peak  [Fig.~4(k)], than in the 1JHei  [Fig.~4(i)].
The observation of a strong excitation
at $(\pi,0)$ in K-edge RIXS~\cite{PhysRevB.81.085124} is also explained by our calculations, which
demonstrate a concentration of DOS at exactly this wave-vector. Again
the one-magnon states that contributes to this peak [Fig.~4(l)] are very different from the 1JHei~[Fig.~4(j)]. 
Thus, our results reveal dramatic differences in the two-magnon continuum, implying that INS, $L_3$ or
$K$-edge RIXS and Raman data must be interpreted using the full quantitative model derived here.\\
Our model also provide insight into the electronic spectra, such as the bare-band dispersion used to extract the self-energy function from ARPES spectra.
A self-consistent Kramers-Kroniger analysis of ARPES experiments on  Bi2212 revealed $t\simeq0.23$~eV~\cite{PhysRevB.71.214513}.
However, with the discovery of a high-energy kink~\cite{PhysRevLett.98.067004} at 0.4~eV in the nodal spectrum (also known as the waterfall feature), 
a similar analysis on optimally doped LSCO~\cite{PhysRevB.78.205103} suggested that $t\simeq0.48$~eV. Our results for LCO ($t=0.492$~eV) support the second scenario.
%
%

In summary we derived an effective spin Hamiltonian valid for any lattice
and any hopping matrix element range. Using spin-wave theory with $1/S$-corrections and
three hopping $t$, $t'$ and $t''$, we obtain accurate quantitative 1bHub\ 
parameter sets for several parent compounds of the high-T$_c$ cuprate 
superconductors. We predict ordered moment, double
occupancy, SW renormalization and 2-magnon spectra.
From the non-interacting two-magnon
$S^{zz}(\bm q,\omega)$ and DOS, we clearly demonstrate the necessity to 
include the extended exchange paths to interpret Raman and K-edge RIXS peaks.
Furthermore, electronic spectra such as ARPES could also be addressed using the same 1bHub\
parameters.

We gratefully acknowledge Headings \textit{et al.}~\cite{PhysRevLett.105.247001}
for sharing their data, J. Chang, F.~Vernay, F.~Mila, T.~A.~T\`oth, B.~Normand 
and M.~E.~Zhitomirsky for fruitful discussions, K. J. Zhou for his help with the RIXS experiment and the 
Swiss NSF and the MaNEP NCCR for support.

\bibliographystyle{apsrev4-1}
\bibliography{references}

\begin{thebibliography}{10}%
\makeatletter
\providecommand \@ifxundefined [1]{%
 \ifx #1\undefined \expandafter \@firstoftwo
 \else \expandafter \@secondoftwo
\fi
}%
\providecommand \@ifnum [1]{%
 \ifnum #1\expandafter \@firstoftwo
 \else \expandafter \@secondoftwo
\fi
}%
\providecommand \enquote [1]{``#1''}%
\providecommand \bibnamefont  [1]{#1}%
\providecommand \bibfnamefont [1]{#1}%
\providecommand \citenamefont [1]{#1}%
\providecommand\href[0]{\@sanitize\@href}%
\providecommand\@href[1]{\endgroup\@@startlink{#1}\endgroup\@@href}%
\providecommand\@@href[1]{#1\@@endlink}%
\providecommand \@sanitize [0]{\begingroup\catcode`\&12\catcode`\#12\relax}%
\@ifxundefined \pdfoutput {\@firstoftwo}{%
 \@ifnum{\z@=\pdfoutput}{\@firstoftwo}{\@secondoftwo}%
}{%
 \providecommand\@@startlink[1]{\leavevmode\special{html:<a href="#1">}}%
 \providecommand\@@endlink[0]{\special{html:</a>}}%
}{%
 \providecommand\@@startlink[1]{%
  \leavevmode
  \pdfstartlink
   attr{/Border[0 0 1 ]/H/I/C[0 1 1]}%
   user{/Subtype/Link/A<</Type/Action/S/URI/URI(#1)>>}%
  \relax
 }%
 \providecommand\@@endlink[0]{\pdfendlink}%
}%
\providecommand \url  [0]{\begingroup\@sanitize \@url }%
\providecommand \@url [1]{\endgroup\@href {#1}{\urlprefix}}%
\providecommand \urlprefix [0]{URL }%
\providecommand \Eprint[0]{\href }%
\@ifxundefined \urlstyle {%
  \providecommand \doi [1]{doi:\discretionary{}{}{}#1}%
}{%
  \providecommand \doi [0]{doi:\discretionary{}{}{}\begingroup
  \urlstyle{rm}\Url }%
}%
\providecommand \doibase [0]{http://dx.doi.org/}%
\providecommand \Doi[1]{\href{\doibase#1}}%
\providecommand \bibAnnote [3]{%
  \BibitemShut{#1}%
  \begin{quotation}\noindent
    \textsc{Key:}\ #2\\\textsc{Annotation:}\ #3%
  \end{quotation}%
}%
\providecommand \bibAnnoteFile [2]{%
  \IfFileExists{#2}{\bibAnnote {#1} {#2} {\input{#2}}}{}%
}%
\providecommand \typeout [0]{\immediate \write \m@ne }%
\providecommand \selectlanguage [0]{\@gobble}%
\providecommand \bibinfo [0]{\@secondoftwo}%
\providecommand \bibfield [0]{\@secondoftwo}%
\providecommand \translation [1]{[#1]}%
\providecommand \BibitemOpen[0]{}%
\providecommand \bibitemStop [0]{}%
\providecommand \bibitemNoStop [0]{.\EOS\space}%
\providecommand \EOS [0]{\spacefactor3000\relax}%
\providecommand \BibitemShut [1]{\csname bibitem#1\endcsname}%
\bibitem{anderson_resonating_1987}%
  \BibitemOpen
  \bibfield{author}{%
  \bibinfo {author} {\bibfnamefont{P.~W.}\ \bibnamefont{{Anderson}}},\ }%
  \bibfield{journal}{%
  \Doi{10.1126/science.235.4793.1196}{\bibinfo {journal} {Science}}\ }%
  \textbf{\bibinfo {volume} {235}},\ \bibinfo {pages} {1196 } (\bibinfo {month}
  {Mar.}\ \bibinfo {year} {1987})%
  \bibAnnoteFile{NoStop}{anderson_resonating_1987}%
\bibitem{macdonald_t/u_1988}%
  \BibitemOpen
  \bibfield{author}{%
  \bibinfo {author} {\bibfnamefont{A.~H.}\ \bibnamefont{{MacDonald}}}
  \emph{et~al.},\ }%
  \bibfield{journal}{%
  \Doi{10.1103/PhysRevB.37.9753}{\bibinfo {journal} {Phys. Rev. B}}\ }%
  \textbf{\bibinfo {volume} {37}},\ \bibinfo {pages} {9753} (\bibinfo {month}
  {Jun.}\ \bibinfo {year} {1988})%
  \bibAnnoteFile{NoStop}{macdonald_t/u_1988}%
\bibitem{coldea_spin_2001}%
  \BibitemOpen
  \bibfield{author}{%
  \bibinfo {author} {\bibfnamefont{R.}~\bibnamefont{Coldea}} \emph{et~al.},\ }%
  \bibfield{journal}{%
  \Doi{10.1103/PhysRevLett.86.5377}{\bibinfo {journal} {Phys. Rev. Let.}}\ }%
  \textbf{\bibinfo {volume} {86}},\ \bibinfo {pages} {5377} (\bibinfo {month}
  {Jun.}\ \bibinfo {year} {2001})%
  \bibAnnoteFile{NoStop}{coldea_spin_2001}%
\bibitem{RevModPhys.75.473}%
  \BibitemOpen
  \bibfield{author}{%
  \bibinfo {author} {\bibfnamefont{A.}~\bibnamefont{Damascelli}}
  \emph{et~al.},\ }%
  \bibfield{journal}{%
  \Doi{10.1103/RevModPhys.75.473}{\bibinfo {journal} {Rev. Mod. Phys.}}\ }%
  \textbf{\bibinfo {volume} {75}},\ \bibinfo {pages} {473} (\bibinfo {month}
  {Apr}\ \bibinfo {year} {2003})%
  \bibAnnoteFile{NoStop}{RevModPhys.75.473}%
\bibitem{delannoy_low-energy_2009}%
  \BibitemOpen
  \bibfield{author}{%
  \bibinfo {author} {\bibfnamefont{J.~P.}\ \bibnamefont{Delannoy}}
  \emph{et~al.},\ }%
  \bibfield{journal}{%
  \Doi{10.1103/PhysRevB.79.235130}{\bibinfo {journal} {Phys. Rev. B}}\ }%
  \textbf{\bibinfo {volume} {79}},\ \bibinfo {pages} {235130} (\bibinfo {month}
  {Jun.}\ \bibinfo {year} {2009})%
  \bibAnnoteFile{NoStop}{delannoy_low-energy_2009}%
\bibitem{PhysRevLett.105.157006}%
  \BibitemOpen
  \bibfield{author}{%
  \bibinfo {author} {\bibnamefont{Guarise}} \emph{et~al.},\ }%
  \bibfield{journal}{%
  \Doi{10.1103/PhysRevLett.105.157006}{\bibinfo {journal} {Phys. Rev. Lett.}}\
  }%
  \textbf{\bibinfo {volume} {105}},\ \bibinfo {pages} {157006} (\bibinfo
  {month} {Oct}\ \bibinfo {year} {2010})%
  \bibAnnoteFile{NoStop}{PhysRevLett.105.157006}%
\bibitem{PhysRevLett.105.247001}%
  \BibitemOpen
  \bibfield{author}{%
  \bibinfo {author} {\bibnamefont{Headings}} \emph{et~al.},\ }%
  \bibfield{journal}{%
  \Doi{10.1103/PhysRevLett.105.247001}{\bibinfo {journal} {Phys. Rev. Lett.}}\
  }%
  \textbf{\bibinfo {volume} {105}},\ \bibinfo {pages} {247001} (\bibinfo
  {month} {Dec}\ \bibinfo {year} {2010})%
  \bibAnnoteFile{NoStop}{PhysRevLett.105.247001}%
\bibitem{ISI:000236516300002}%
  \BibitemOpen
  \bibfield{author}{%
  \bibinfo {author} {\bibfnamefont{P.~A.}\ \bibnamefont{Lee}} \emph{et~al.},\
  }%
  \bibfield{journal}{%
  \bibinfo {journal} {Rev. of Mod. Phys.}\ }%
  \textbf{\bibinfo {volume} {78}},\ \bibinfo {pages} {17} (\bibinfo {month}
  {Jan}\ \bibinfo {year} {2006})%
  \bibAnnoteFile{NoStop}{ISI:000236516300002}%
\bibitem{Oguchi1960}%
  \BibitemOpen
  \bibfield{author}{%
  \bibinfo {author} {\bibfnamefont{T.}~\bibnamefont{Oguchi}},\ }%
  \bibfield{journal}{%
  \Doi{10.1103/PhysRev.117.117}{\bibinfo {journal} {Phys. Rev.}}\ }%
  \textbf{\bibinfo {volume} {117}},\ \bibinfo {pages} {117} (\bibinfo {month}
  {Jan}\ \bibinfo {year} {1960})%
  \bibAnnoteFile{NoStop}{Oguchi1960}%
\bibitem{ISI:000086788300001}%
  \BibitemOpen
  \bibfield{author}{%
  \bibinfo {author} {\bibfnamefont{T.}~\bibnamefont{Tohyama}}\ and\ \bibinfo
  {author} {\bibfnamefont{S.}~\bibnamefont{Maekawa}},\ }%
  \bibfield{journal}{%
  \bibinfo {journal} {Supercond. Sci. Technol.}\ }%
  \textbf{\bibinfo {volume} {13}},\ \bibinfo {pages} {R17} (\bibinfo {month}
  {Apr}\ \bibinfo {year} {2000})%
  \bibAnnoteFile{NoStop}{ISI:000086788300001}%
\bibitem{PhysRevB.69.094515}%
  \BibitemOpen
  \bibfield{author}{%
  \bibinfo {author} {\bibfnamefont{Y.-D.}\ \bibnamefont{Chuang}}
  \emph{et~al.},\ }%
  \bibfield{journal}{%
  \Doi{10.1103/PhysRevB.69.094515}{\bibinfo {journal} {Phys. Rev. B}}\ }%
  \textbf{\bibinfo {volume} {69}},\ \bibinfo {pages} {094515} (\bibinfo {month}
  {Mar}\ \bibinfo {year} {2004})%
  \bibAnnoteFile{NoStop}{PhysRevB.69.094515}%
\bibitem{PhysRevB.72.224511}%
  \BibitemOpen
  \bibfield{author}{%
  \bibinfo {author} {\bibfnamefont{J.}~\bibnamefont{Lorenzana}} \emph{et~al.},\
  }%
  \bibfield{journal}{%
  \Doi{10.1103/PhysRevB.72.224511}{\bibinfo {journal} {Phys. Rev. B}}\ }%
  \textbf{\bibinfo {volume} {72}},\ \bibinfo {pages} {224511} (\bibinfo {month}
  {Dec}\ \bibinfo {year} {2005})%
  \bibAnnoteFile{NoStop}{PhysRevB.72.224511}%
\bibitem{Hasan09062000}%
  \BibitemOpen
  \bibfield{author}{%
  \bibinfo {author} {\bibfnamefont{M.~Z.}\ \bibnamefont{Hasan}} \emph{et~al.},\
  }%
  \bibfield{journal}{%
  \Doi{10.1126/science.288.5472.1811}{\bibinfo {journal} {Science}}\ }%
  \textbf{\bibinfo {volume} {288}},\ \bibinfo {pages} {1811} (\bibinfo {year}
  {2000})%
  \bibAnnoteFile{NoStop}{Hasan09062000}%
\bibitem{PhysRevB.81.085124}%
  \BibitemOpen
  \bibfield{author}{%
  \bibinfo {author} {\bibfnamefont{D.~S.}\ \bibnamefont{Ellis}} \emph{et~al.},\
  }%
  \bibfield{journal}{%
  \Doi{10.1103/PhysRevB.81.085124}{\bibinfo {journal} {Phys. Rev. B}}\ }%
  \textbf{\bibinfo {volume} {81}},\ \bibinfo {pages} {085124} (\bibinfo {month}
  {Feb}\ \bibinfo {year} {2010})%
  \bibAnnoteFile{NoStop}{PhysRevB.81.085124}%
\bibitem{Canali1993}%
  \BibitemOpen
  \bibfield{author}{%
  \bibinfo {author} {\bibfnamefont{C.~M.}\ \bibnamefont{Canali}}\ and\ \bibinfo
  {author} {\bibfnamefont{M.}~\bibnamefont{Wallin}},\ }%
  \bibfield{journal}{%
  \Doi{10.1103/PhysRevB.48.3264}{\bibinfo {journal} {Phys. Rev. B}}\ }%
  \textbf{\bibinfo {volume} {48}},\ \bibinfo {pages} {3264} (\bibinfo {month}
  {Aug}\ \bibinfo {year} {1993})%
  \bibAnnoteFile{NoStop}{Canali1993}%
\bibitem{Canali1992}%
  \BibitemOpen
  \bibfield{author}{%
  \bibinfo {author} {\bibfnamefont{C.~M.}\ \bibnamefont{Canali}}\ and\ \bibinfo
  {author} {\bibfnamefont{S.~M.}\ \bibnamefont{Girvin}},\ }%
  \bibfield{journal}{%
  \Doi{10.1103/PhysRevB.45.7127}{\bibinfo {journal} {Phys. Rev. B}}\ }%
  \textbf{\bibinfo {volume} {45}},\ \bibinfo {pages} {7127} (\bibinfo {month}
  {Apr}\ \bibinfo {year} {1992})%
  \bibAnnoteFile{NoStop}{Canali1992}%
\bibitem{Nagao2007}%
  \BibitemOpen
  \bibfield{author}{%
  \bibinfo {author} {\bibfnamefont{T.}~\bibnamefont{Nagao}}\ and\ \bibinfo
  {author} {\bibfnamefont{J.-I.}\ \bibnamefont{Igarashi}},\ }%
  \bibfield{journal}{%
  \Doi{10.1103/PhysRevB.75.214414}{\bibinfo {journal} {Phys. Rev. B}}\ }%
  \textbf{\bibinfo {volume} {75}},\ \bibinfo {pages} {214414} (\bibinfo {month}
  {Jun}\ \bibinfo {year} {2007})%
  \bibAnnoteFile{NoStop}{Nagao2007}%
\bibitem{christensen_quantum_2007}%
  \BibitemOpen
  \bibfield{author}{%
  \bibinfo {author} {\bibfnamefont{N.~B.}\ \bibnamefont{Christensen}}
  \emph{et~al.},\ }%
  \bibfield{journal}{%
  \Doi{10.1073/pnas.0703293104}{\bibinfo {journal} {P. Natl. Acad. Sci. USA}}\
  }%
  \textbf{\bibinfo {volume} {104}},\ \bibinfo {pages} {15264 } (\bibinfo {year}
  {2007})%
  \bibAnnoteFile{NoStop}{christensen_quantum_2007}%
\bibitem{Tsyrulin2010}%
  \BibitemOpen
  \bibfield{author}{%
  \bibinfo {author} {\bibfnamefont{N.}~\bibnamefont{Tsyrulin}}, \bibinfo
  {author} {\bibfnamefont{F.}~\bibnamefont{Xiao}}, \bibinfo {author}
  {\bibfnamefont{A.}~\bibnamefont{Schneidewind}}, \bibinfo {author}
  {\bibfnamefont{P.}~\bibnamefont{Link}}, \bibinfo {author}
  {\bibfnamefont{H.~M.}\ \bibnamefont{Ronnow}}, \bibinfo {author}
  {\bibfnamefont{J.}~\bibnamefont{Gavilano}}, \bibinfo {author}
  {\bibfnamefont{C.~P.}\ \bibnamefont{Landee}}, \bibinfo {author}
  {\bibfnamefont{M.~M.}\ \bibnamefont{Turnbull}},\ and\ \bibinfo {author}
  {\bibfnamefont{M.}~\bibnamefont{Kenzelmann}},\ }%
  \bibfield{journal}{%
  \bibinfo {journal} {Phys. Rev. B}\ }%
  \textbf{\bibinfo {volume} {81}} (\bibinfo {year} {2010})%
  \bibAnnoteFile{NoStop}{Tsyrulin2010}%
\bibitem{PhysRevLett.87.037202}%
  \BibitemOpen
  \bibfield{author}{%
  \bibinfo {author} {\bibfnamefont{H.~M.}\ \bibnamefont{R\o{}nnow}}, \bibinfo
  {author} {\bibfnamefont{D.~F.}\ \bibnamefont{McMorrow}}, \bibinfo {author}
  {\bibfnamefont{R.}~\bibnamefont{Coldea}}, \bibinfo {author}
  {\bibfnamefont{A.}~\bibnamefont{Harrison}}, \bibinfo {author}
  {\bibfnamefont{I.~D.}\ \bibnamefont{Youngson}}, \bibinfo {author}
  {\bibfnamefont{T.~G.}\ \bibnamefont{Perring}}, \bibinfo {author}
  {\bibfnamefont{G.}~\bibnamefont{Aeppli}}, \bibinfo {author}
  {\bibfnamefont{O.}~\bibnamefont{Sylju\aa{}sen}}, \bibinfo {author}
  {\bibfnamefont{K.}~\bibnamefont{Lefmann}},\ and\ \bibinfo {author}
  {\bibfnamefont{C.}~\bibnamefont{Rischel}},\ }%
  \bibfield{journal}{%
  \Doi{10.1103/PhysRevLett.87.037202}{\bibinfo {journal} {Phys. Rev. Lett.}}\
  }%
  \textbf{\bibinfo {volume} {87}},\ \bibinfo {pages} {037202} (\bibinfo {month}
  {Jun}\ \bibinfo {year} {2001})%
  \bibAnnoteFile{NoStop}{PhysRevLett.87.037202}%
\bibitem{Christensen2004}%
  \BibitemOpen
  \bibfield{author}{%
  \bibinfo {author} {\bibfnamefont{N.~B.}\ \bibnamefont{Christensen}}, \bibinfo
  {author} {\bibfnamefont{D.~F.}\ \bibnamefont{McMorrow}}, \bibinfo {author}
  {\bibfnamefont{H.~M.}\ \bibnamefont{R\o{}nnow}}, \bibinfo {author}
  {\bibfnamefont{A.}~\bibnamefont{Harrison}}, \bibinfo {author}
  {\bibfnamefont{T.~G.}\ \bibnamefont{Perring}},\ and\ \bibinfo {author}
  {\bibfnamefont{R.}~\bibnamefont{Coldea}},\ }%
  \bibfield{journal}{%
  \bibinfo {journal} {J. Mag. Mag. Mat.}\ }%
  \textbf{\bibinfo {volume} {272}},\ \bibinfo {pages} {896} (\bibinfo {year}
  {2004})%
  \bibAnnoteFile{NoStop}{Christensen2004}%
\bibitem{PhysRevB.53.R11930}%
  \BibitemOpen
  \bibfield{author}{%
  \bibinfo {author} {\bibfnamefont{G.}~\bibnamefont{Blumberg}} \emph{et~al.},\
  }%
  \bibfield{journal}{%
  \Doi{10.1103/PhysRevB.53.R11930}{\bibinfo {journal} {Phys. Rev. B}}\ }%
  \textbf{\bibinfo {volume} {53}},\ \bibinfo {pages} {R11930} (\bibinfo {month}
  {May}\ \bibinfo {year} {1996})%
  \bibAnnoteFile{NoStop}{PhysRevB.53.R11930}%
\bibitem{PhysRevB.71.214513}%
  \BibitemOpen
  \bibfield{author}{%
  \bibinfo {author} {\bibfnamefont{A.~A.}\ \bibnamefont{Kordyuk}}, \bibinfo
  {author} {\bibfnamefont{S.~V.}\ \bibnamefont{Borisenko}}, \bibinfo {author}
  {\bibfnamefont{A.}~\bibnamefont{Koitzsch}}, \bibinfo {author}
  {\bibfnamefont{J.}~\bibnamefont{Fink}}, \bibinfo {author}
  {\bibfnamefont{M.}~\bibnamefont{Knupfer}},\ and\ \bibinfo {author}
  {\bibfnamefont{H.}~\bibnamefont{Berger}},\ }%
  \bibfield{journal}{%
  \Doi{10.1103/PhysRevB.71.214513}{\bibinfo {journal} {Phys. Rev. B}}\ }%
  \textbf{\bibinfo {volume} {71}},\ \bibinfo {pages} {214513} (\bibinfo {month}
  {Jun}\ \bibinfo {year} {2005})%
  \bibAnnoteFile{NoStop}{PhysRevB.71.214513}%
\bibitem{PhysRevLett.98.067004}%
  \BibitemOpen
  \bibfield{author}{%
  \bibinfo {author} {\bibfnamefont{J.}~\bibnamefont{Graf}}, \bibinfo {author}
  {\bibfnamefont{G.-H.}\ \bibnamefont{Gweon}}, \bibinfo {author}
  {\bibfnamefont{K.}~\bibnamefont{McElroy}}, \bibinfo {author}
  {\bibfnamefont{S.~Y.}\ \bibnamefont{Zhou}}, \bibinfo {author}
  {\bibfnamefont{C.}~\bibnamefont{Jozwiak}}, \bibinfo {author}
  {\bibfnamefont{E.}~\bibnamefont{Rotenberg}}, \bibinfo {author}
  {\bibfnamefont{A.}~\bibnamefont{Bill}}, \bibinfo {author}
  {\bibfnamefont{T.}~\bibnamefont{Sasagawa}}, \bibinfo {author}
  {\bibfnamefont{H.}~\bibnamefont{Eisaki}}, \bibinfo {author}
  {\bibfnamefont{S.}~\bibnamefont{Uchida}}, \bibinfo {author}
  {\bibfnamefont{H.}~\bibnamefont{Takagi}}, \bibinfo {author}
  {\bibfnamefont{D.-H.}\ \bibnamefont{Lee}},\ and\ \bibinfo {author}
  {\bibfnamefont{A.}~\bibnamefont{Lanzara}},\ }%
  \bibfield{journal}{%
  \Doi{10.1103/PhysRevLett.98.067004}{\bibinfo {journal} {Phys. Rev. Lett.}}\
  }%
  \textbf{\bibinfo {volume} {98}},\ \bibinfo {pages} {067004} (\bibinfo {month}
  {Feb}\ \bibinfo {year} {2007})%
  \bibAnnoteFile{NoStop}{PhysRevLett.98.067004}%
\bibitem{PhysRevB.78.205103}%
  \BibitemOpen
  \bibfield{author}{%
  \bibinfo {author} {\bibfnamefont{J.}~\bibnamefont{Chang}} \emph{et~al.},\ }%
  \bibfield{journal}{%
  \Doi{10.1103/PhysRevB.78.205103}{\bibinfo {journal} {Phys. Rev. B}}\ }%
  \textbf{\bibinfo {volume} {78}},\ \bibinfo {pages} {205103} (\bibinfo {month}
  {Nov}\ \bibinfo {year} {2008})%
  \bibAnnoteFile{NoStop}{PhysRevB.78.205103}%
\end{thebibliography}%
\end{document}